\documentclass[twocolumn,showpacs,preprintnumbers,amsmath,amssymb]{revtex4}
\usepackage{graphicx,epsfig}
\usepackage{dcolumn}
\usepackage{bm}
\begin{document}

\title{Relationship between Entropic Bottleneck in Free Energy Landscape, 
Nonexponential Relaxation and Fragility of Glass-Forming Liquids}
\author{Dwaipayan Chakrabarti and Biman Bagchi\footnote[2] 
{For correspondence: bbagchi@sscu.iisc.ernet.in}}
\affiliation{Solid State and Structural Chemistry Unit, Indian Institute 
of Science, Bangalore 560012, India}
\date{\today}
\begin{abstract}
A mesoscopic model is proposed to explain the anomalous dynamics in a 
supercooled liquid as its glass transition temperature is approached 
from above. The model is based on the assumption of $\beta$ organized 
$\alpha$ process, with the requirement of coherent excitation of a minimum 
critical number $N_{c}$ of $\beta$ processes in the surroundings of
a total $N_{\beta}$. Numerical evaluation of the model shows that 
the growth in this critical number in the background of a modest $N_{\beta}$
can lead to a severe entropic bottleneck and slow down the dynamics to an 
extent observed near real glass transition. The fragility of the glass-forming 
liquid is shown to be correlated with the growth of the ratio 
$\gamma ( = N_{c}/N_{\beta})$.
\end{abstract}

\maketitle 

Relaxation in a viscous liquid above but near its glass transition 
temperature $T_{g}$ has been a subject of immense attention in recent time 
\cite{Angel-Ngai-McKenna-McMillan-Martin, Ediger-Angell-Nagel, 
Debenedetti-Stillinger, Angell-JNCS, Stillinger-Science-1995, SDS, 
Mohanty}. The dramatic increase in the characteristic relaxation time 
$\tau$ (which varies as much as $15$ orders of magnitude for fragile liquids 
\cite{Angell-JNCS, Sastry-Nature}) is most frequently described by the 
Vogel-Fulcher-Tammann (VFT) equation  \cite{VFT} 
\begin{equation}
\tau = A_{VFT}~exp[B/(T - T_{0})],
\end{equation}
where $A_{VFT}$, $B$ and $T_{0}$ are found to remain constant over a range of
$4-5$ orders of magnitude variation of $\tau$. $T_{0}$, known as the Vogel 
temperature, is typically $30-50$ K below $T_{g}$. $T_{0} = 0$ corresponds to 
the Arrhenious behavior, characterizing the strong liquid limit. The fragile 
behavior is further marked by nonexponentiality in the relaxation functions, 
which can be described by a stretched exponential or Kohlrausch-Williams-Watts 
function 
\begin{equation} 
\phi(t) = exp[-(t/\tau)^\beta],
\end{equation}
where $\beta$, $0 < \beta < 1$, is the stretching exponent. For a typical 
fragile glass-former, $\beta$ generally decreases from near $1$ at high 
temperatures to below $0.5$ close to $T_{g}$ with a display of nearly monotonic 
temperature dependence \cite{Angel-Ngai-McKenna-McMillan-Martin}. 
Experimental and computational studies suggest that the 
stretched exponential relaxation may owe its origin to the growth of 
spatially heterogeneous relaxing domains, where each individual exhibits a 
nearly exponential relaxation with a relaxation time that varies 
significantly among the individuals 
\cite{Ediger-ARPC-Sillescu-JNC-Richert-JPMC, heterogeneity}. Such heterogeneous 
domains have been found experimentally to span $2-3$ nm 
\cite{Cicerone-Blackburn-Ediger-JCP, Tracht-PRL}. 
However, the origin of this modest size of the heterogeneous domains is 
not clearly understood at present. A consistent description of all the above 
aspects of dynamics in the supercooled regime has remained a major scientific 
challenge over many decades. 

Several theories suggest the appearance of a growing length scale as the 
glass transition is approached from above 
\cite{Xia-Wolynes-PNAS}. The celebrated Adam-Gibbs (AG) theory \cite{Adam-Gibbs} 
attempts to explain the temperature dependence of relaxation time in terms 
of a temperature variation of the size $z$ of the cooperatively rearranging 
region (CRR); the lower limit $z^{\star}$ is "shown" to follow the relation 
$z^{\star} = N_{A}s_{c}^{\star}/S_{c}$, where $s_{c}^{\star}$ is the 
critical configurational entropy of a CRR corresponding to 
$z^{\star}$, $S_{c}$ is the molar configurational entropy of the 
macroscopic sample, and $N_{A}$ is the Avogadro constant. According to 
AG theory, the sluggishness near $T_{g}$ is due to the scarcity of the 
number of configurations available to a CRR, which is reflected in the 
increasingly smaller value of $S_{c}$; the resultant rapid increase in the 
value of $z^{\star}$ leads to the divergence of the relaxation time $\tau$. 
However, experimental \cite{Cicerone-Blackburn-Ediger-JCP} and 
computer simulation \cite{Dasgupta} studies have failed to find a 
convincing evidence of a growing length scale near $T_{g}$. Nevertheless, 
AG theory of entropy crisis provides a useful conceptual framework.

Perhaps the most successful quantitative theory of relaxation phenomena 
in the supercooled liquid state is the mode-coupling theory (MCT) 
\cite{Bengtzelius, Gotze-Sjogren}. It is known to become inadequate at 
low temperatures below $T_{c}$, called the MCT critical temperature. This is
presumably because of the prevalence of the thermally activated hopping 
\cite{Bengtzelius, Angell-JPCS} below $T_{c}$ (unaccounted for in MCT) between 
the adjacent minima of the energy landscape 
\cite{Goldstein, Stillinger-Science-1995}. Recent 
computer simulation studies have revealed that hopping is a highly 
cooperative phenomenon, promoted by many body fluctuations 
\cite{hopping}. Computer simulations have further shown that large amplitude 
hopping of a tagged particle is often preceded by somewhat larger than normal, 
still small amplitude, but collective motion, of its neighbors. A similar 
picture of $\beta$ and $\alpha$ processes was earlier proposed by Stillinger, 
where the $\beta$ process corresponds to transitions between inherent 
structures (IS) within the meta-basin while $\alpha$ process corresponds to 
transitions between deep meta-basins \cite{Stillinger-Science-1995}.
Motivated by these findings, we here present a mesoscopic model of relaxation
in supercooled liquids ($\sim$ below $T_{c}$), where an $\alpha$ process is 
{\it promoted by coherent excitations of a minimum number of $\beta$ processes 
within a CRR.} The $\beta$ processes are assumed, as a first approximation, 
to occur independently. The requirement of coherence is that a given minimum 
number $N_{c}$ among the total number $N_{\beta}$ of $\beta$ processes 
must be in the excited state during a small interval for an $\alpha$ process 
(i.e., transition out of a meta-basin) to occur. In our model, a CRR is 
characterized by an $N_{\beta}$ number of identical noninteracting 
two-level systems (TLSs). Each of these TLSs transits back and forth at 
equilibrium at temperature T between its two levels, labeled $0$ (ground) and 
$1$ (excited). The two levels are separated by an energy $\epsilon$. The 
waiting time before a transition can occur from the level 
$i (= 0, 1)$ is random, but is drawn from a Poissonian probability density 
function given by
\begin{equation} 
\psi_{i}(t) = \frac{1}{\tau_{i}} exp(-t/\tau_{i}), ~~~~~~~~~~~~~~i = 0, 1,
\end{equation}
where $\tau_{i}$ is the average time of stay in the level $ i $. If $p_{i}$ 
denotes the canonical equilibrium probability of the level $i$ being occupied, 
detailed balance gives the following relation
\begin{equation}
K = \frac{p_{1}}{p_{0}} = \frac{\tau_{1}}{\tau_{0}},
\end{equation}
where $ K $ is the equilibrium constant for the two levels. We define a 
variable $\zeta_{j}(t),~(j = 1, 2, ....., N_{\beta})$, which takes on a value 
$0$ if at the given instant of time $t$ the level $0$ of the two-level system
$j$ is occupied and $1$ if otherwise. $\zeta_{j}(t)$ is thus an 
occupation variable. The control variable $Q(t)$ is defined as
\begin{equation}
Q(t) = \displaystyle \sum_{j=1}^{N_{\beta}} \zeta_{j}(t).
\end{equation}
Q(t), which serves as an order parameter, is a stochastic variable in the 
discrete integer space $[0, N_{\beta}]$ and carries information of the 
excitation prevailing in the CRR at time $t$. The rate of $\alpha$ relaxation 
depends crucially on Q. For simplicity, we assume here that the relaxation 
occurs with unit probability at the instant $Q$ reaches $N_{c}$, an integer 
greater than the most probable value of $Q$, for the first time. This 
restriction can be removed, but only at the expense of the analytical solution 
presented below.

We solve the model for the average relaxation time $\tau$, for a
a given pair of $N_{c}$ and $N_{\beta}$, by using the method of mean first 
passage time \cite{Gardiner}. The probability that the stochastic variable 
$Q$ takes on a value $l$ at time $t$, $P(l; t)$, satisfies the following 
master equation
\begin{eqnarray}
\frac {dP(l;t)}{dt} &=& [(N_{\beta} - l + 1)/\tau_{0}]P(l - 1;t) \nonumber 
\\ &+& [(l + 1)/\tau_{1}]P(l + 1;t) \nonumber \\ &-& 
[(N_{\beta} - l)/\tau_{0}]P(l;t) - (l/\tau_{1})P(l;t).
\end{eqnarray}
The mean first passage time $\tau (l)$,  which is the mean time elapsed before 
the stochastic variable $Q$ (starting from its initial value $l \leq N_{c} - 1$) 
reaches $ N_{c} $ for the first time, satisfies the following equation related 
to the backward master equation:
\begin{equation}
[(N_{\beta} - l)/\tau_{0}][\tau(l + 1) - \tau(l)] + (l/\tau_{1})[\tau(l - 1) - 
\tau(l)] = - 1,
\end{equation} 
subject to an absorbing boundary condition at $ Q = N_{c} $, $\tau(N_{c}) = 0$, 
and a reflecting boundary condition at $ Q = 0 $, $ \tau(-1) = \tau(0) $. 
We solve Eq. (7) to obtain $\tau (l)$ as a sum over hypergeometric 
functions 
$F(a,b;c;z)$
\begin{eqnarray}
\tau(l) &=& \tau_{0} (1 + 1/K)^{N_{\beta}} \nonumber \\ &~& \displaystyle 
\sum_{n = l}^{N_{c}-1} 
\frac {F(N_{\beta} + 1,N_{\beta} - n;N_{\beta} - n + 1;-1/K)}{N_{\beta} - n}.
\end{eqnarray}
Eq. (8) shows that when the two states are of the same energy, that is, $K = 1$, 
even then the relaxation slows down significantly. This is purely an entropic 
effect -- a nice example of entropic bottleneck \cite{Zwanzig}. When there is 
an energy bias against the excited state (state 1), the bottleneck becomes more 
severe, as shown below.

\begin{figure}[tb]
\epsfig{file=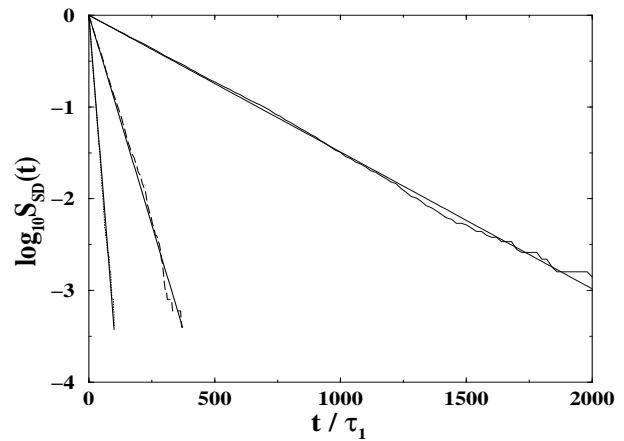,height=8cm,width=6cm,angle=-90}
\caption{Semi-log plot of the single domain relaxation function $S_{SD}(t)$ 
versus $t$ at a given $T$. The figure is based on the data obtained by 
simulating the model for single domain relaxation with $\epsilon = k_{B}T$, 
$k_{B}$ being the Boltzmann constant, and 
$N_{\beta} = 10$ for  $N_{c} = 6, 7$ and $8$ (dotted line, long dashed line, 
and dot-dashed line, respectively). The solid lines are the straight line fits.}
\end{figure}
\begin{figure}[tb]
\epsfig{file=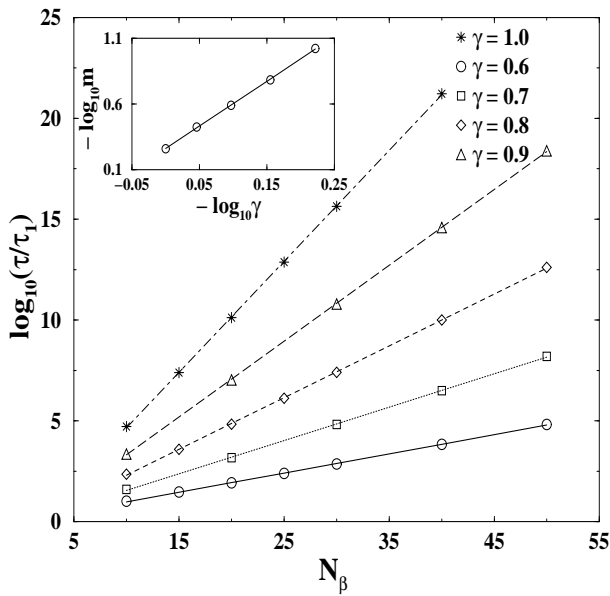,height=8cm,width=8cm,angle=-90}
\caption{Dependence of the single domain average relaxation time $\tau_{SD}$ on 
$N_{\beta}$ at a given temperature $T$ with $\epsilon / (k_{B}T) = 1$ for 
different fixed values of $\gamma$. The inset shows the $\gamma$ dependence of 
the slope $m$ of the semi-log plot of the scaled $\tau_{SD}$ versus 
$N_{\beta}$.}
\end{figure}
\begin{figure}[tb]
\epsfig{file=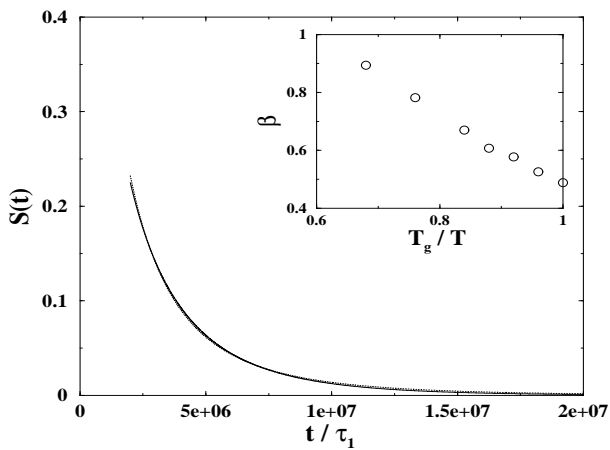,height=8cm,width=6cm,angle=-90}
\caption{The long time behavior of the averaged relaxation function $S(t)$. The 
solid line is a fit to the stretched exponential function with 
$\tau = 1.100 \times 10^{6} \tau_{1}$ and $\beta = 0.670$. The inset shows the 
temperature dependence of the stretching exponent $\beta$.}
\end{figure}
\begin{figure}[tb]
\epsfig{file=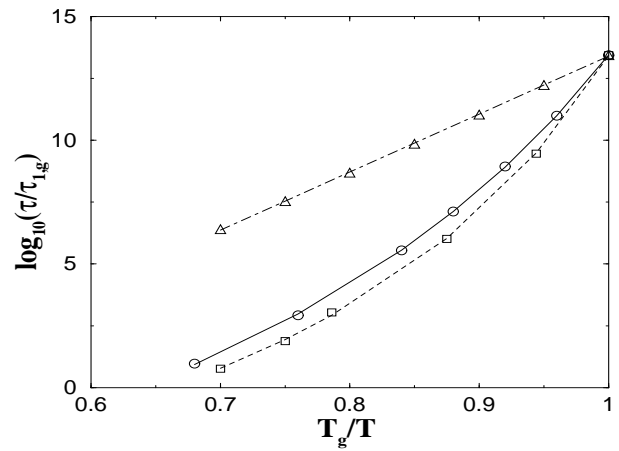,height=8cm,width=6cm,angle=-90}
\caption{Arrhenious plot showing the scaled characteristic relaxation time
$\tau$ as a function of reduced inverse temperature $T_{g}/T$. The two sets of 
data, marked with circles and squares, for $\tau$ are obtained from the 
stretched exponential fit of the long time behavior of the average relaxation 
function $S(t)$. The circles correspond to the data set obtained from a linear 
variation of $\gamma$ from $0.6$ to $1.0$ as $T_{g}/T$ is increased from $0.68$ 
to $1.0$ with $N_{\beta}$ held fixed at $20$. The squares denote the data set 
corresponding to a linear growth of $\gamma$ with $T_{g}/T$, both rising from 
$0.70$ to $1.0$, with $N_{c}$ and $N_{\beta}$ varied together, the latter from 
$10$ to $20$. The solid and the dotted lines, which are the corresponding fits 
to the VFT form with $T_{0} = 0.689~T_{g}$ and $0.789~T_{g}$, respectively, 
illustrate fragile liquid behavior. The data set, marked with triangles, are 
obtained with $\gamma$, $N_{\beta}$, and $\epsilon$ all held 
constant: $\gamma = 1.0$, $N_{\beta} = 40$, and $\epsilon = k_{B}T_{m}$. The 
data are scaled so as to have the same $\tau$ value at $T_{g}$ as for the other 
two sets. With no distribution of $\epsilon$, the relaxation is exponential with 
the time constant $\tau$. The straight line fit 
corresponding to the Arrhenious behavior is typical of a strong liquid. Here, 
for the sake of simplicity, numerical evaluation is done by taking 
$\tau_{SD} = \tau (Q_{mp})$, where $Q_{mp}$ is the most probable value of $Q$; 
this involves negligible error.}
\end{figure} 

The relaxation of the relevant time correlation functions, such as stress and 
density, is assumed to be caused by the $\alpha$ relaxation. The time 
dependence of the $\alpha$ relaxation can be quantified by the survival 
probability correlation function $S_{SD}$, of the initial meta-basin state. 
This is obtained by first simulating the model for single domain relaxation. 
The parameter $\gamma$, defined as $\gamma = N_{c}/N_{\beta}$, provides a
measure of the free energy barrier to relaxation having both the energy and the 
entropy contributions. Fig. (1) plots the time dependence of the single domain 
relaxation function $S_{SD}(t)$ in the logarithmic scale at a given temperature 
for three values of $N_{c}$. The straight line fits show a single exponential 
decay of $S_{SD}(t)$. The time constants obtained are identical to those 
obtained from our analytical expression, given by Eq. (8). The single domain 
relaxation is found to slow down considerably, as expected, with increasing 
$\gamma$. Fig. (2) illustrates an exponential dependence of the single domain 
average relaxation time $\tau_{SD}$ on $N_{\beta}$ at fixed $\gamma$ and 
$T$; $\tau_{SD}$  is a weighted average of $\tau (l)$ over the initial 
distribution: 
$\tau_{SD} = \displaystyle \sum_{l = 0}^{N_{c}-1} P(l;0)\tau(l)$.
The dependence, characterized by the slope of the semi-log plot of the scaled 
$\tau_{SD}$ versus $N_{\beta}$, becomes stronger as $\gamma$ increases. The 
inset suggests a power law behavior of the slope $m$, $m \sim \gamma ^{\nu}$ 
with $\nu \simeq 3.4$. Note that $\tau_{SD}$ is also the mean waiting time.

In a heterogeneous environment within a bulk sample, a fluid-like region can be 
characterized by having, at a given time, on the average, a relatively large 
number of $\beta$ processes in the excited state. The reverse is true for 
solid-like regions. Therefore, a Gaussian distribution of $\epsilon$ among CRRs 
can incorporate the existence of heterogeneous domains into the model 
(one can also assume an exponential distribution). 
The present calculation takes the mean $<\epsilon>$ of the distribution to be 
unity and the standard deviation $\sigma = 0.05$ in the units of $k_{B}T_{m}$, 
$T_{m}$ being the melting temperature. $T_{m}$ is also used to define $T_{g}$: 
$T_{g} = 2T_{m}/3$. Furthermore, the model assumes $N_{c}$ to grow as the 
reduced inverse temperature $T_{g}/T$ increases until $N_{c}$ reaches 
$N_{\beta}$ at $T_{g}$. The Gaussian distribution 
of $\epsilon$ results in a continuous distribution of $\tau_{SD}$. The average 
relaxation function $S(t)$ is calculated from 
$ S(t) = \int \limits^{\infty}_0 d\tau_{SD} g(\tau_{SD}) exp(-t/\tau_{SD})$, 
where $g(\tau_{SD})$ is the probability density function. The long time 
behavior of $S(t)$ fits well to the stretched exponential function as shown in 
Fig. (3). The inset shows a monotonic decrease of the stretching exponent 
$\beta$ with $T_{g}$ approached from above, as indeed observed experimentally.
The study of the temperature dependence of the characteristic relaxation time 
$\tau$ requires time to be scaled by the same unit at all temperatures. 
$\tau_{1}$ at $T_{g}$, which we denote by $\tau_{1,g}$, is chosen for that. The 
scaling needs, for a transition state theory (TST) calculation, an input to 
$\epsilon^{\ddagger}$, the energy barrier to the transition from the level 
$0$ to $1$, which is taken to be $4~k_{B}T_{m}$ and held constant. Fig. (4) 
shows the $T_{g}$ scaled Arrhenious plot of the characteristic relaxation time. 
The fit to the VFT equation is good. The growth of $\gamma$ with $T$ approaching
$T_{g}$ from above determines the fragility, as illustrated in Fig. (4). Strong 
liquid limit is obtained with $\gamma$ held fixed at $1$. Strong liquids are 
spatially (and, dynamically) correlated even at high temperatures. This 
justifies the choice of a higher $\gamma$ and $N_{\beta}$ values in the strong
liquid limit. 

We have also calculated the waiting time distribution $W(\tau)$ (WTD) 
\cite{Doliwa-Heuer} for this $\alpha$ process. Even when the distribution of 
energy gap in the meta basin is Gaussian, WTD is non-Gaussian with a stretching 
at long $\tau$. The waiting time distribution gets modified if the distribution 
of energy gap ($\epsilon$) is exponential -- the most affected region is 
obviously the small $\tau$ limit.
 
The present model can be considered as a representative in a simple form of a 
class of wider, more general models. An immediate generalization will be to 
include the correlations among the $\beta$ processes within a CRR. Note that 
the present model explains the slow down of relaxation in fragile liquids as 
result of an entropic bottleneck superimposed on the energy constraint and does 
not require any diverging length scale. The model predicts that the observed 
$13$ orders of magnitude increase in relaxation time originates from the 
combined effect of energy and entropy, and it is not possible to separate the 
two effects. The model is based on a requirement of dynamical correlation 
between the $\beta$ and $\alpha$ processes. This correlation itself depends on 
the time scale of separation between the two. In the language of energy 
landscape \cite{Stillinger-Science-1995, Heuer, Doliwa-Heuer}, 
the $\beta$ transitions assumed here occur within the super-structures 
(meta-basins) of a deep minimum and they promote transitions between two deep 
minima -- that is, among the meta basins. While the model rests on dwindling 
entropy as the temperature is lowered, it does not invoke any thermodynamic 
phase transition. Most notably, only a modest growth in the size of CRR, 
represented here by the size of $N_{\beta}$, is required to capture the 
experimentally observed slow down. 

We thank Professor S. Sastry, Dr. S. Bhattacharyya, Mr. R. Murarka and Mr. A. 
Mukherjee for many helpful discussions. This work was supported in parts by 
grants from CSIR, DAE and DST, India. DC acknowledges Fellowship from 
University Grants Commission (UGC), India.

\end{document}